\documentclass{ecai} 

\usepackage{latexsym}
\usepackage{amssymb}
\usepackage{amsmath}
\usepackage{amsthm}
\usepackage{multirow}
\usepackage{booktabs}
\usepackage{enumitem}
\usepackage{graphicx}
\usepackage{color}
\usepackage{array}
\usepackage{makecell}
\usepackage{subcaption}

\usepackage{todonotes}
\newcommand{\BibTeX}{B\kern-.05em{\sc i\kern-.025em b}\kern-.08em\TeX}

\begin{document}


\begin{frontmatter}


\paperid{8888} 


\title{The Heterogeneous Multi-Agent Challenge}


\author[A]{\fnms{Charles}~\snm{Dansereau}\thanks{Corresponding Author. Email: charles.dansereau@thalesgroup.com.}}
\author[A]{\fnms{Junior Samuel}~\snm{Lopez Yepez}}
\author[A]{\fnms{Karthik}~\snm{Soma}}
\author[A]{\fnms{Antoine}~\snm{Fagette}}

\address[A]{THALES, cortAIx Labs Canada}


\begin{abstract}
Multi-Agent Reinforcement Learning (MARL) is a growing research area which gained significant traction in recent years, extending Deep RL applications to a much wider range of problems. A particularly challenging class of problems in this domain is Heterogeneous Multi-Agent Reinforcement Learning (HeMARL), where agents with different sensors, resources, or capabilities must cooperate based on local information. The large number of real-world situations involving heterogeneous agents makes it an attractive research area, yet underexplored, as most MARL research focuses on homogeneous agents (e.g., a swarm of identical robots). In MARL and single-agent RL, standardized environments such as ALE and SMAC have allowed to establish recognized benchmarks to measure progress. However, there is a clear lack of such standardized testbed for cooperative HeMARL. As a result, new research in this field often uses simple environments, where most algorithms perform near optimally, or uses weakly heterogeneous MARL environments. 

In this paper, we address this gap by proposing the Heterogeneous Multi-Agent Challenge (HeMAC),\footnotemark \: a new benchmarking environment based on the PettingZoo standard. HeMAC features a suite of challenges across multiple scenarios, offering varied and controllable complexity and agent heterogeneity. Our results show that while agents using advanced algorithms such as MAPPO excel in simpler cooperative tasks, their performance declines as heterogeneity increases, with IPPO outperforming them in highly diverse scenarios. QMIX struggles significantly under these conditions due to its assumptions of shared action values and agent homogeneity.  These findings demonstrate HeMAC’s value as a rigorous testbed for evaluating MARL algorithms in heterogeneous settings, and emphasize the need for further research in this field to handle complexity and heterogeneity effectively.

\end{abstract}
\end{frontmatter}

\footnotetext[1]{{Code is available at: \url{https://github.com/ThalesGroup/hemac}}}

\section{Introduction}

Deep Reinforcement Learning (DRL) has demonstrated remarkable performance in various domains, from robotics to strategic decision-making \cite{openai_hand_manipulation, alphago, alphastar}. While many early successes in RL were in single-agent settings, a wide range of real-world problems is inherently multi-agent in nature. Examples include autonomous vehicle coordination, distributed sensor networks, robotic swarms, energy grid management, or network traffic routing. Multi-Agent Reinforcement Learning (MARL) provides a natural framework for tackling such problems by allowing multiple learning agents to interact within a shared environment. However, MARL introduces significant challenges, including decentralization constraints, exponentially growing joint action spaces, and the complexities of learning coordinated behavior in dynamic and partially observable settings.
\begin{figure}[t!]
\centering
\includegraphics[width=\columnwidth]{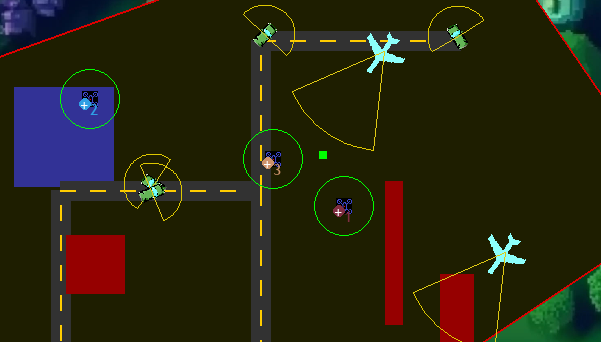}
\vspace{-2mm}
\caption{An example scenario in HeMAC. A team of heterogeneous agents coordinate to find and retrieve a target (green square).}
\label{fig:sharing}
\vspace{5mm}
\end{figure}

Among the various MARL paradigms, heterogeneous MARL (HeMARL), where agents differ in their action spaces, observation capabilities, or objectives, is of particular interest. Many real-world scenarios involve heterogeneous agents, such as teams of Unmanned Aerial Vehicles (UAV) and Unmanned Ground Vehicles (UGV), multi-role autonomous fleets, and mixed-agent strategic simulations. Unlike homogeneous MARL, where all agents share similar capabilities, HeMARL introduces additional complexities related to role differentiation, specialized coordination, and adaptive strategy formation. Despite these challenges, research in HeMARL has been growing, with various approaches emerging to tackle decentralized coordination and policy learning for heterogeneous teams \cite{bettini_heterogeneous_2023, liu_heterogeneous_2022, heterogeneous_traffic}. Recent work has proposed Heterogeneous-Agent Reinforcement learning (HARL), and adapted state-of-the-art algorithms such as PPO, TD3, and TRPO for HeMARL problems, proposing HAPPO, HATD3, and HATRPO \cite{zhong_heterogeneous_agent_2024}.

However, a key limitation in the field is the lack of standardized benchmarks and testbeds for HeMARL research. Although single agent RL has benefited from widely adopted benchmarking environments such as the Arcade Learning Environment (ALE) \cite{ALE} and MuJoCo \cite{mujoco}, and homogeneous MARL has been advanced by environments such as the StarCraft Multi-Agent Challenge (SMAC and SMACv2) \cite{SMAC, SMACv2} and the Google Research Football Environment (GRF) \cite{GRF}, a standard for heterogeneous multi-agent settings is missing. Existing multi-agent environments often assume homogeneous agent capabilities \cite{Hanabi, GRF}, or contain heterogeneity but do not scale beyond simple grid-world scenarios \cite{MPE, pettingzoo}, making it difficult to systematically evaluate HeMARL algorithms in complex, high-dimensional settings. This has led to inconsistent evaluation methodologies, where researchers frequently develop one-off environments that may be biased toward specific algorithms or fail to capture the full scope of HeMARL challenges.

To address this gap, we introduce the Heterogeneous Multi-Agent Challenge (HeMAC), a new benchmark designed specifically for heterogeneous multi-agent reinforcement learning. Following OpenAI's Gym interface \cite{gym_openai_2016}, we base HeMAC on the PettingZoo framework, a multi-agent version of the newer Gymnasium standard \cite{towers_gymnasium_2024}. More specifically, HeMAC uses PettingZoo's Agent-Environment Cycle (AEC) API \cite{pettingzoo}, to provide a structured environment where agents with distinct characteristics must cooperate to achieve their goals. By enforcing decentralized execution and enabling flexible agent heterogeneity, HeMAC presents a rich set of challenges, including role specialization, asymmetric information sharing, and multi-agent credit assignment. The benchmark is designed to facilitate rigorous and reproducible HeMARL research by offering well-defined tasks, standardized evaluation metrics, and diverse scenarios that reflect the complexities of heterogeneous systems in the real world. We include results in various scenarios of HeMAC using QMIX \cite{rashid_qmix_2018}, IPPO \cite{ippo}, MAPPO \cite{mappo} and baseline algorithms, and challenge the community to make progress on multi-agent problems with highly heterogeneous settings.

\section{Related Work}
A substantial amount of work has been done to propose challenging environments to test MARL methods. However, only very few studies focus on cases with heterogeneity between agents in terms of observation spaces, action spaces, or even capabilities and dynamics.

Table \ref{het_envs} presents a list of the main environments with heterogeneous agents. The Butterfly environment, released with the PettingZoo multi-agent framework, proposes some tasks with heterogeneous agents that can scale in difficulty with the number of agents \cite{pettingzoo}. However, the heterogeneity of the agents is limited in that the agents have similar capabilities (e.g., Cooperative Pong) or observation spaces (e.g., Knight Archers Zombies -KAZ-). In addition, the agents require minimal coordination, as the environment lacks dynamics for asymmetric interactions between the agents. The Prospector task was initially proposed as a complex challenge requiring higher coordination, but was removed from the Butterfly environment and is no longer supported by PettingZoo.

In the Multiple-Particle Environment (MPE)\cite{MPE}, some of the tasks include different observation spaces between agents, such as World-Comm. In this predator-prey task, a team of agents collaborates with a leader to catch a faster prey. Other tasks also include different action spaces: in the Speaker-Listener, a speaker agent tries to guide a listener agent to a landmark with communication actions. Speaker-Listener and World-Comm were successfully used in \cite{MPE} to validate the adaptation of the actor-critic method to mixed cooperative-competitive environments. Nevertheless, neither of these tasks offers complex interactions or scalability in difficulty with respect to scenarios or number of agents, reducing the appeal of the environment to be a challenging testbed for novel HeMARL techniques. 

More complex MARL environments have been presented that contain heterogeneity, such as SMAC (and later SMACv2) \cite{SMAC, SMACv2}, which is a well established benchmark in MARL. In SMAC, the components of heterogeneity come from the different units that are used in the multitude of scenarios proposed. For example, some units cannot attack enemy units and instead "heal" ally units. SMAC has the advantage of offering complex scenarios, which makes it challenging and a good benchmark to compare novel MARL methods. However, the heterogeneity in its agents is severely limited. Aside from the previously mentioned variations in capabilities, the agents possess identical observation and action spaces and operate within the same environment. The only differences between the agents arise from a few parameters, such as health and speed, or attacking versus healing, the shape of the observation and action spaces being the same. Methods developed and tested using this environment risk facing severe difficulties when applied to complex teams of varied platforms with vastly different observation spaces, action spaces, and capabilities across each agent. For example, QMIX, a widely used multi-agent DRL algorithm, assumes homogeneous observation and action spaces, making it unsuitable for direct application in these types of challenges \cite{rashid_qmix_2018}.

HeMAC addresses these limitations by introducing a new class of MARL environments explicitly designed with agents with distinct observation spaces, action spaces, and transition dynamics. This level of heterogeneity not only gives access to an under-researched class of real-world multi-agent systems (e.g., mixed teams of drones, maritime vehicles, or ground robots), but also allows researchers to systematically study the scalability and adaptability of MARL methods in the presence of deep agent asymmetry, or help measure progress of new HeMARL techniques, such as HARL, with a more comprehensive set of heterogeneous problems.

\begin{table}[]
\caption{Overview of the existing MARL environments with cooperative heterogeneous agents. Heterogeneity can be present in the form of different observation spaces, action spaces or capabilities.}
\label{het_envs}
\resizebox{\columnwidth}{!}{%
\begin{tabular}{@{}m{2.0cm}m{4.4cm}clll@{}}
\toprule
\toprule
\multicolumn{1}{c}{\multirow{3}{*}{Environment}} &
  \multicolumn{1}{c}{\multirow{3}{*}{Task summary}} &
  \multicolumn{1}{c}{\multirow{3}{*}{\begin{tabular}[c]{@{}c@{}}Difficulty /\\ scalability\end{tabular}}} &
  \multicolumn{3}{c}{Heterogeneity between agents} \\ \cmidrule(l){4-6} 
\multicolumn{1}{c}{} &
  \multicolumn{1}{c}{} &
  \multicolumn{1}{c}{} &
  \multicolumn{1}{c}{\begin{tabular}[c]{@{}c@{}}Observation\\ Space\end{tabular}} &
  \multicolumn{1}{c}{\begin{tabular}[c]{@{}c@{}}Action\\ Space\end{tabular}} &
  \multicolumn{1}{c}{\begin{tabular}[c]{@{}c@{}}Dynamics /\\ capabilities\end{tabular}} \\ \midrule \midrule
\makecell{Speaker Listener \\(MPE)} &
  A speaker tries to guide a listener to move to a target point. &
  Very simple &
  Different &
  Different &
  Different \\  \midrule
\makecell{World Comm \\(MPE)} &
  Predator-prey environment. A team of agents tries to catch faster agents. The team consists of pursuers and a leader pursuer that has a global observation and needs to communicate. &
  Simple &
  \makecell{Partially \\ Different} &
  \makecell{Partially \\ Different} &
  Different \\  \midrule
Rover Tower &
  Multi-agent version of Speaker Listener. Multiple towers try to guide their respective rover to a goal. &
  Simple &
  Different &
  Different &
  Different \\  \midrule
\makecell{Cooperative Pong \\(Butterfly)} &
  Two agents controlling asymmetric paddles cooperate to play pong. &
  Very simple &
  Identical &
  Identical &
  \makecell{Partially \\ Different} \\  \midrule
\makecell{Knights Archers \\Zombies (Butterfly)} &
  Two types of agents (archers and knights) cooperate to eliminate zombies. Archers shoot arrows while knights swing their mace to kill zombies. &
  Moderate &
  Identical &
  Identical &
  \makecell{Partially \\ Different} \\  \midrule
\makecell{Prospector \\(Butterfly)} &
  Banker and prospector agents work together to collect gold, share it to one another and deposit it in banks. \textbf{Note}: Prospector has been removed from PettingZoo and is no longer maintained. &
  Moderate &
  Different &
  Different &
  Different \\  \midrule
\makecell{SMAC \\ (some scenarios)} &
  Units with different roles (e.g., healers, melee, ranged) cooperate to defeat an enemy army. &
  \makecell{Moderate \\to complex} &
  Identical &
  \makecell{Partially \\ Different} &
  \makecell{Partially \\ Different} \\ \bottomrule
\bottomrule
\end{tabular}%
}
\end{table}

\section{Heterogeneous Multi-Agent Reinforcement Learning}

\paragraph{DEC-POMDP.} A partially observable, cooperative multi-agent reinforcement learning task with heterogeneous agents can be described as a general Decentralized Partially Observable Markov Decision Process (Dec-POMDP). This extends the classical Dec-POMDP formalism to accommodate differences in agents' action spaces, observation spaces, and transition dynamics. Formally, it is defined as a tuple $(n, S, \{A_i\}_{i=1}^n, T, \{O_i\}_{i=1}^n, O, R, \gamma)$, where $n$ is the number of agents, $S$ is the set of environment states, $A_i$ is the set of actions available to agent $i$, $T : S \times A_1 \times \dots \times A_n \rightarrow \Delta(S)$ is the transition function, $O_i$ is the set of observations for agent $i$, $O : S \times A_1 \times \dots \times A_n \rightarrow \Delta(O_1 \times \dots \times O_n)$ is the observation function, $R : S \times A_1 \times \dots \times A_n \rightarrow \mathbb{R}$ is the shared reward function, and $\gamma \in [0, 1]$ is the discount factor. At each timestep $t$, agent $i$ selects an action $a_i \in A_i$, forming a joint action $\mathbf{a} = (a_1, \dots, a_n)$. The environment transitions from state $s$ to $s'$ with probability $T(s, \mathbf{a})$, emits a reward $r = R(s, \mathbf{a})$, and provides each agent with an individual observation $o_i \in O_i$ sampled from $O(s, \mathbf{a})$. Each agent builds its own trajectory $\tau_i \in (O_i \times A_i)^*$, and aims to learn a policy $\pi_i : \tau_i \rightarrow \Delta(A_i)$ that maximises the expected cumulative reward $\mathbb{E}\left[\sum_t \gamma^t r_t\right]$ over time.
\begin{figure*}[th]
    \centering
    \begin{subfigure}{0.285\textwidth}
        \centering
        \includegraphics[width=\textwidth]{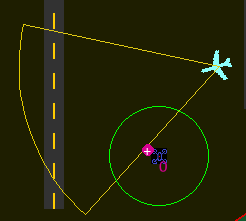}
        \caption{Simple Fleet}
        \label{fig:simple_fleet}
    \end{subfigure}
    \hfill
    \begin{subfigure}{0.375\textwidth}
        \centering
        \includegraphics[width=\textwidth]{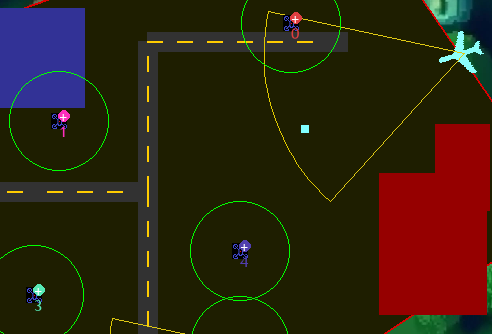}
        \caption{Fleet}
        \label{fig:fleet}
    \end{subfigure}
    \hfill
    \begin{subfigure}{0.31\textwidth}
        \centering
        \includegraphics[width=\textwidth]{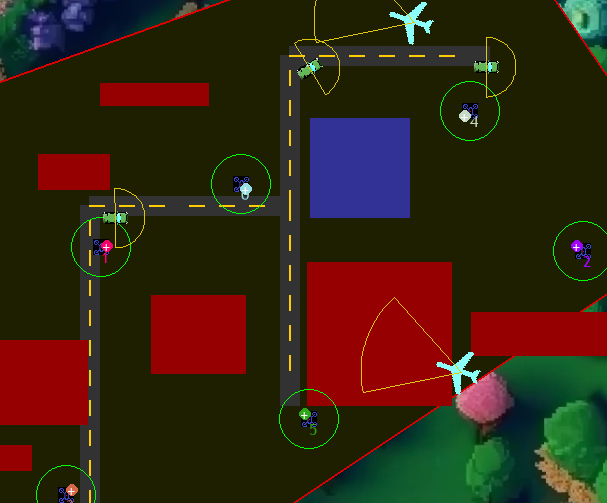}
        \caption{Complex Fleet}
        \label{fig:complex_fleet}
    \end{subfigure}
    \vspace{3mm}
    \caption{Screenshots of HeMAC Challenges involving heterogeneous agents. From left to right: \textbf{Simple Fleet} — a single Quadcopter and Observer. \textbf{Fleet} — multiple Quadcopters now operate with a recharge area (blue square) and buildings to avoid (red). A target is detected by an Observer (cyan). \textbf{Complex Fleet} — a dense environment with multiple obstacles, Quadcopters, Observers, and Provisioners navigating a road network.}
    \label{fig:hemac_challenges}
    \vspace{3mm}
\end{figure*}
\paragraph{Observation and Action Space Padding.} 
In order to apply standard multi-agent reinforcement learning algorithms to heterogeneous settings, a common practice is to homogenize agents' observation and action spaces through padding. This involves extending all observation and action vectors to a common dimensionality, often by zero-padding shorter vectors or using placeholder values. Additionally, continuous action spaces are sometimes discretized, or discrete actions are embedded into continuous spaces to enable uniform policy architectures. However, such approaches introduce several limitations: padding increases the dimensionality of the input and output spaces, which can significantly slow down learning and lead to inefficient policy representations. Furthermore, mappings between continuous and discrete spaces may fail to preserve the expressiveness of the original action space or result in incomplete coverage of the action domain, ultimately leading to suboptimal performance in complex environments. Hence, HeMAC aims to drive research in HeMARL techniques that can be applied without these practices by deliberately introducing mixed (continuous and discrete) observation and action spaces with differing dimensions. Still, the observation and action spaces are offered in both formats and can be padded to benchmark against classical MARL algorithms that might require homogeneous, padded spaces.

\paragraph{Agent-Environment Cycle API.}
The proposed environment is implemented using the Agent-Environment Cycle (AEC) API provided by the PettingZoo library. This API formalizes multi-agent interaction as a turn-based sequence, where agents act one at a time in a predefined or environment-defined order. We refer readers to \cite{pettingzoo} for more details. The AEC abstraction simplifies the design of complex multi-agent environments by enforcing clear control flow and agent scheduling, making it easier to implement asymmetric interactions and handle agents with heterogeneous action and observation spaces. Furthermore, it ensures compatibility with a broad range of reinforcement learning libraries and enables deterministic environment rollouts, which are useful for debugging and reproducibility.

\section{HeMAC}
HeMAC is a 2D physics-based environment built on the PettingZoo framework, which is the multi-agent version of the Gymnasium standard. In HeMAC, a team of autonomous agents with different capabilities needs to coordinate to find and reach moving targets in a randomly generated map. Three types of agents (Quadcopter, Observer, and Provisioner, described hereafter) offer different capabilities, and the team needs to learn how to collaborate and make use of its specialized agents to reach targets in the most efficient way. The challenge comes from two components: role specialization, induced by capability differences, and effective coordination between agents, which is required for optimal performance. The task is inspired by the NP-hard multi-depot vehicle routing problem \cite{MDVR} in order to offer a rich and complex problem to benchmark MARL techniques. Note that more types of agent can be created with various capabilities, observations, and action spaces, as one wishes to complexify the challenge even more, but we initially propose three. To ease the creation of new agents, we define a base agent with the minimal requirements to be directly compatible with HeMAC, from which the proposed agents all derive. 

Three challenges have been crafted with increasing levels of complexity and heterogeneity between agents, offering a range of difficulties to make it a convenient environment for developing and testing algorithms. In each of the challenges, multiple scenarios are offered and can be expanded as needed with various numbers of agents, increasing or reducing the level of coordination required. A summary of the challenges with their proposed scenarios is presented in Table \ref{challenges}. Additionally, one can create new scenarios with only one type of agent or in single agent settings, making HeMAC compatible with traditional single agent and homogeneous multi-agent methods, if needed. HeMAC provides together partial observability, different observation and action spaces, different capabilities and complex interactions between the agents.

\paragraph{Challenge 1: \textit{Simple Fleet}.} In Simple Fleet, two types of agents coordinate in a simplified version of the problem: the team needs to reach a single moving target as many times as possible in a fixed number of timesteps. When reached, another target is randomly generated in the area. The first category of agents, Quadcopters, is characterized by low-altitude flight and high agility, enabling them to effectively reach their designated targets. In contrast, the second category, Observers, resembles fixed-wing unmanned aerial vehicles (UAVs) and operates at significantly higher speeds and altitudes. However, Observers lack the capability to approach the targets closely enough to reach them.

Observers have a much larger observation field of view and need to communicate with the Quadcopters to guide them towards the targets. We propose multiple scenarios with a variable number of each type of agents to gradually increase complexity. An example scenario is shown in Figure \ref{fig:simple_fleet}.

\paragraph{Challenge 2: \textit{Fleet}.} In Fleet, we extend the problem to multi-target search, increase the heterogeneity between the agents, and add complexity in the environment. The Quadcopters now have a limited energy capacity and frequently need to go back to a charging point to continue the mission, or they risk running out of battery. In contrast, Observers do not have energy constraints and should learn different strategies to make use of their different endurance. In addition, obstacles (e.g., buildings) are generated and placed randomly in each episode. Those obstacles require Quadcopters to navigate around them, whereas Observers operate at high enough altitudes to ignore collisions with those obstacles. However, they must remain within a specific range of a building to successfully transmit communications to their teammates. An example scenario is shown in Figure \ref{fig:fleet}.

\paragraph{Challenge 3: \textit{Complex Fleet}.} Complex Fleet aims to be a challenging testbed, containing a high level of heterogeneity to require HeMARL techniques to fully take into account the different capabilities of the agents, their different observation and action spaces, and how they interact with each other. We introduce a third type of agent, the Provisioner, that is an autonomous ground vehicle that navigates a road network to help the team reach multiple targets more efficiently. In this challenge, Quadcopters not only have limited energy but also limited carry capacity. As an additional requirement, Quadcopters must bring back the targets they reach to a gathering point but can do so only one target at a time.

Provisioners are agents with greater autonomy and capacity. They can recharge the Quadcopters and retrieve the targets they reach or those brought back by Quadcopters, allowing the members of the team to continue their mission. They need to learn to coordinate with the Observers and the Quadcopters to strategically position themselves to maximize the efficiency of the team in terms of bringing back targets. In addition to the obstacles, a road network is generated in each episode, and Provisioners are restricted to navigate within it. This means that each agent will interact and navigate in the world with different dynamics: the Quadcopters perform 2D holonomic navigation with obstacle avoidance, the Observers instead need to navigate with 2D airplane-like dynamics, and the Provisioners need to navigate along a road network that can be represented by a graph. An example scenario of Complex Fleet is shown in Figure \ref{fig:complex_fleet}.

\paragraph{Representation of HeMARL problems} The rationale behind HeMAC's design is to target key challenges in HeMARL: coordinating agents with different capabilities under partial observability. Each challenge introduces a different layer of heterogeneity, ranging from differences in sensing and control (Simple Fleet), to energy constraints and asymmetric communication (Fleet), to multi-modal mobility and resource transfer (Complex Fleet). This progression allows systematic benchmarking of algorithm scalability and robustness. Although HeMAC does not cover every problem that might be encountered in the HeMARL field, we selected agent types and tasks that reflect real-world scenarios (e.g. drone-UGV coordination or logistics under resource constraints) and the design of the environment aims to be modular and extensible, supporting community-driven growth. Moreover, HeMAC currently focuses on collaborative tasks only. Competitive and mixed-motive HeMARL remains an important area for future exploration.

\begin{figure}[h]
\centering
\includegraphics[width=\columnwidth]{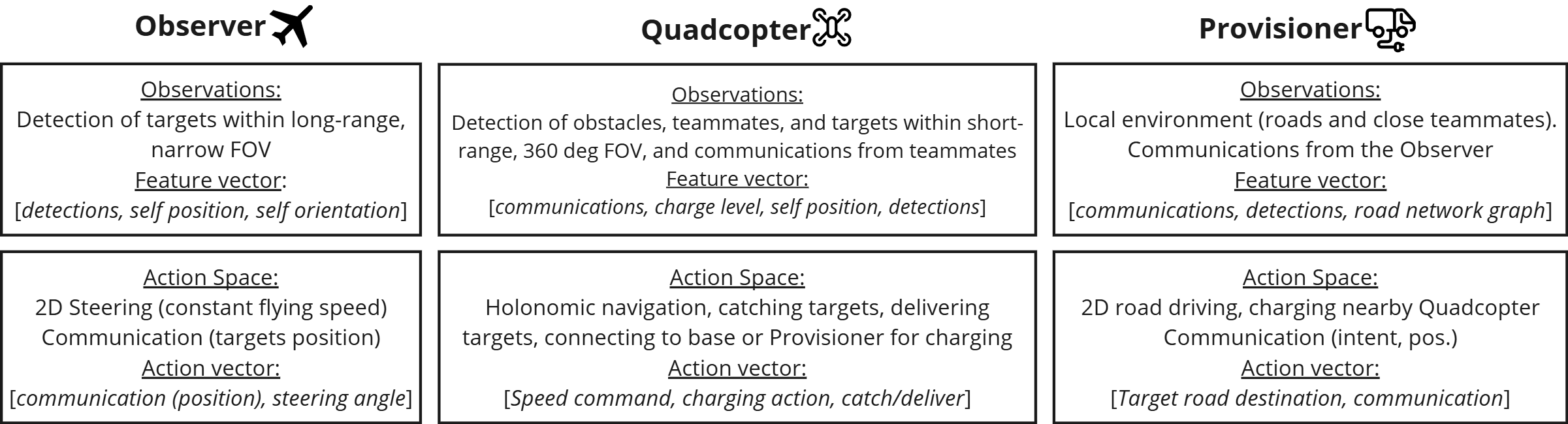}
\caption{Observation and action spaces of the different types of agents.}
\vspace{5mm}
\label{fig:spaces}
\end{figure}

\begin{table}[h]
\caption{Summary of HeMAC Challenges}
\label{challenges}
\resizebox{\columnwidth}{!}{%
\begin{tabular}{@{}clp{4.2cm}c@{}} 
\toprule
\textbf{Name} & \textbf{Agents} & \textbf{Challenge} & \textbf{Scenarios} \\ \midrule
\multirow{3}{*}{Simple Fleet} &
  \multirow{3}{*}{\begin{tabular}[l]{@{}l@{}}Quadcopters\\ Observers\end{tabular}} &
  \multirow{3}{*}{\begin{tabular}[c]{@{}p{4.4cm}@{}}Reach and interact with a target (e.g., find and drop water to a survivor after a natural disaster).\end{tabular}} &
  1q1o \\
              &                 &                                        & 3q1o               \\
              &                 &                                        & 5q2o               \\ \midrule
\multirow{7}{*}{Fleet} &
  \multirow{7}{*}{\begin{tabular}[l]{@{}l@{}}Quadcopters\\ Observers\end{tabular}} &
  \multirow{7}{*}{\begin{tabular}[c]{@{}p{4.4cm}@{}}Reach and interact with multiple targets scattered in a cluttered environment, with limited Quadcopter energy capacity and limited Observers communication range (e.g., find and drop water to multiple survivors after a natural disaster in an urban environment).\end{tabular}} & \\
  &                 &                                        & \\ 
              &                 &                                        & 3q1o \\
              &                 &                                        & 10q3o\\
              &                 &                                        & 20q5o \\
              &                 &                                        & \\ 
        &                 &                                        & \\ \midrule
\multirow{8}{*}{Complex Fleet} &
  \multirow{8}{*}{\begin{tabular}[l]{@{}l@{}}Quadcopters\\ Observers\\ Provisioners\end{tabular}} &
  \multirow{8}{*}{\begin{tabular}[c]{@{}p{4.4cm}@{}}Reach and rescue multiple targets scattered in a cluttered environment, with limited Quadcopter energy and carry capacity, limited Observers communication range, and leveraging Provisioners constrained to a road network (e.g., find multiple survivors after a natural disaster in an urban environment and rescue them to a safe location).\end{tabular}} & \\
              &                 &                                        & \\
              &                 &                                        & 3q1o1p \\
              &                 &                                        & 5q2o1p             \\
              &                 &                                        & 5q1o2p             \\
              &                 &                                        & 10q2o2p            \\
              &                 &                                        & 20q3o5p  \\ 
              &                 &                                        & \\
              &                 &                                        & \\
              \bottomrule
\end{tabular}}
\end{table}
\paragraph{Observation and action spaces.} At each timestep, the agents receive local observations from their field of view (FOV) detecting targets, obstacles, and other agents, as well as information from their respective capabilities, such as communications, battery level or current capacity. Hence, the feature vector is different for each type of agent. The limited detection range from the Field-of-View (FoV) of the agents makes the environment partially observable from their standpoint, and heterogeneous since each type of agents has different FoV range and angular aperture. For example, in our proposed challenges, Quadcopters have a 360-degree sensor with a rather short range while Observers are equipped with a front-facing camera featuring a limited angular aperture but a longer range. The observation space of each agent is summarized in Figure \ref{fig:spaces}.

The global state, only available during centralized training, contains information about all the agents and targets. It includes their position relative to the origin (bottom left of the game area), paired with an encoding of agent types. Moreover, it contains the last actions of all agents and information about the scenario such as the number of obstacles, the position of the base, and the road network graph.

Two action space versions, discrete and continuous, are available for all the agents to offer more flexibility for existing MARL and new HeMARL methods. Similarly to the observation space, the action space of each type of agent is defined according to its capabilities and the dynamics by which it interacts with the environment. For example, the Quadcopter can move in any direction of the horizontal plane, whereas the Observer navigates with steering actions. The action space of each agent is summarized in Figure \ref{fig:spaces}.

\section{Results}
\begin{figure*}[th]
    \centering
    \includegraphics[width=\textwidth]{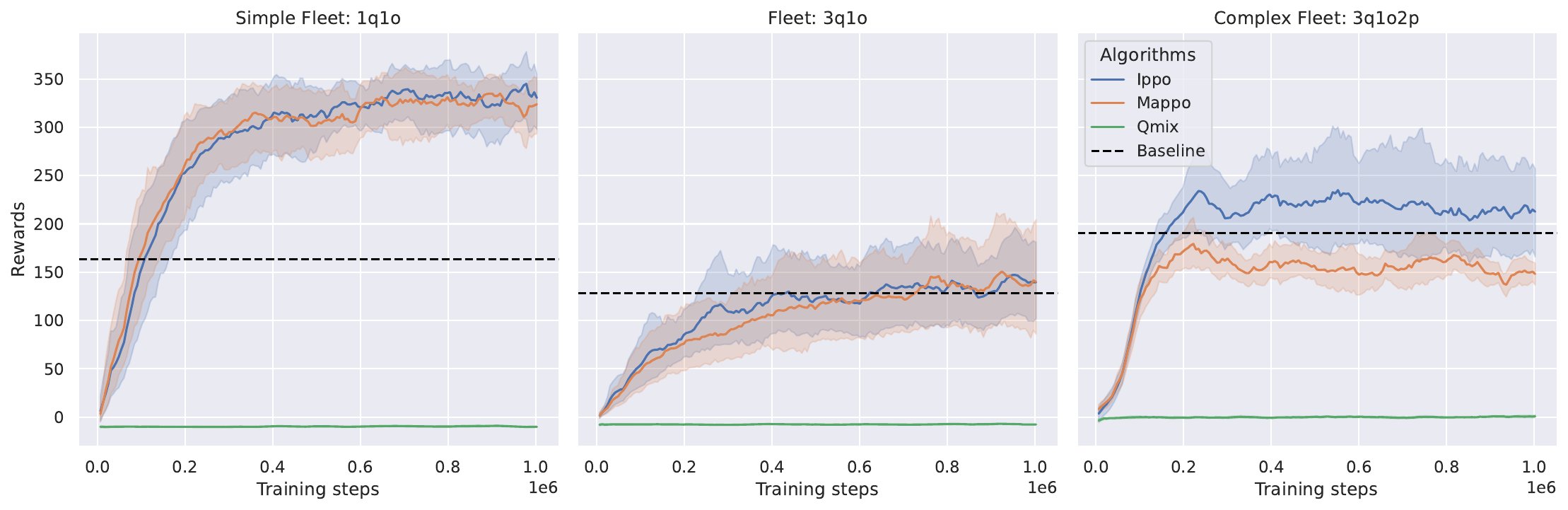}
    \vspace{3mm}
    \caption{Training results of IPPO, MAPPO and QMIX on three scenarios from Simple Fleet, Fleet, and Complex Fleet, respectively. The heuristic performance is shown as a black dotted line.}
    \label{fig:results}
    \vspace{3mm}
\end{figure*}
In this section, we present results on a selection of HeMAC scenarios to demonstrate the suitability of HeMAC as a testbed for the benchmark of MARL algorithms. We assess the performance of current state-of-the-art MARL methods, which in turn illustrates the need for novel HeMARL techniques.

\paragraph{Experiment setup.}
The evaluation method is similar to other work that has proposed standardized environments \cite{SMAC, SMACv2, pettingzoo}. We evaluate multiple state-of-the-art Centralized-Training Decentralized-Execution (CTDE) methods, as well as independent learning techniques for a more complete comparison. We use BenchMARL~\cite{bettini_benchmarl_2024} to benchmark with various algorithms such as IPPO, MAPPO, and QMIX, as it supports the use of heterogeneous action and observation spaces. Note that some algorithms, such as QMIX, assume homogeneous agents and needed to be adapted as they are not directly applicable to heterogeneous settings. For these methods, we padded the observation and actions spaces of the agents to make them compatible (same shape and size). The agent architecture and training details are summarized in Table \ref{tab:agent_details}. Each experiment is repeated with 10 independent runs, which take between 1 to 2 hours for 1 million timesteps on a laptop with an Intel i7-13700K CPU and an NVIDIA RTX 2000 ADA GPU.

\begin{table}[th]
\centering
\caption{Agent architectures and training details.}
\label{tab:agent_details}
\begin{tabular}{ll}
\toprule
\textbf{Component} & \textbf{Description} \\
\midrule
\textbf{Policy Architecture} & 2-layer MLP, 256 units per layer, Tanh activation \\
\textbf{Value Function} & 2-layer MLP, 256 units, Tanh activation\\
\textbf{Optimizer} & Adam, learning rate 0.00005 \\
\textbf{Batch Size} & 1024 \\
\textbf{gradient clipping} & max gradient norm 5 \\
\textbf{Discount Factor} $\gamma$ & 0.9 \\
\textbf{Exploration} & $\epsilon$-greedy decay from 0.8 to 0.01 \\
\textbf{Training Steps} & 1 million timesteps \\
\bottomrule
\end{tabular}
\end{table}

All experiments use the default shaped rewards of HeMAC as the evaluation metric. At each timestep $t$, agent receive shared global rewards $r_t$ according to the the following: 
\begin{equation}
  r_t =
    \begin{cases}
      10 & \text{for reaching a target.} \\
      25 & \text{for retrieving a target (Complex Fleet only).} \\
      0 & \text{otherwise.}
    \end{cases}      
\end{equation}

Additionally, the agents receive individual rewards depending on their type. For instance, Quadcopters receive a penalty of -20 points for colliding with an obstacle, and Observers receive a penalty of -20 points for flying too far from the map.

\paragraph{Experiment results.}
Figure \ref{fig:results} shows the average reward per episode gained by the different algorithms in the easiest scenario of all 3 challenges. We also plot the performance of a heuristic, which is a combination of 3 simple rule-based algorithms, each designed for a specific type of agent. The Quadcopters follow the Observers' communications with a simple collision avoidance strategy, while both the Observers and Provisioners perform a simple exploration strategy that makes sure they do not get out of bounds. The heuristic for each agent is described in Table \ref{tab:heuristics}.
\begin{table}[h]
\centering
\caption{Heuristic policies used for each agent type.}
\label{tab:heuristics}
\begin{tabular}{>{\centering\arraybackslash}m{1.7cm}>{\arraybackslash}m{5.5cm}}
\hline
\textbf{Agent Type} & \textbf{Heuristic Policy} \\
\hline
Quadcopter & Navigate toward the position communicated by Observers, using a potential fields method \cite{apf} for obstacle avoidance. \\
\hline
Observer & Follow a circular patrol trajectory around the map, maintaining proximity to buildings to enable communication of observed target positions. \\
\hline
Provisioner & Drive randomly between intersections on the road network, recharging Quadcopters when they require it. \\
\hline
\end{tabular}
\end{table}

The results show that both IPPO and MAPPO notably outperform the heuristic for Simple Fleet. However, IPPO and MAPPO do not perform significantly higher than the heuristic in the harder scenarios, showing that the increase in complexity and heterogeneity makes it difficult for current algorithms to overcome even simple heuristics in highly heterogeneous settings. In fact, QMIX has the overall worse performance, possibly because of its shared action value and homogeneous agent assumption. As shown in Figure \ref{fig:results}, QMIX performs poorly even in the simplest scenario. Finally, the shortcomings of current MARL methods are shown further in the complex fleet scenario, where MAPPO, normally expected to perform better in multi-agent settings \cite{papoudakis2021benchmarkmarl, mappo}, is performing worse than IPPO and the heuristic when the heterogeneity between agents is the highest. Note that the higher rewards gained by the methods in complex fleet compared to fleet is explained by the additional rewards when retrieving a target. 

\begin{figure}[h]
\centering
\includegraphics[width=0.8\columnwidth]{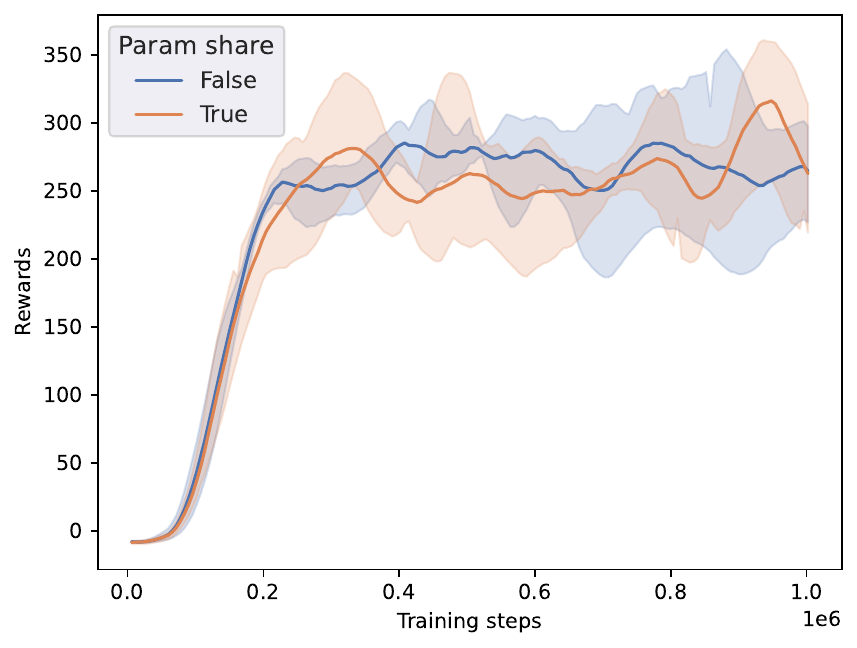}
\caption{Impact of parameter sharing on IPPO performance.}
\label{fig:sharing}
\vspace{5mm}
\end{figure}

Additionally, we benchmark the IPPO algorithm with and without parameter sharing in Figure \ref{fig:sharing}. Parameter sharing is restricted to agents of the same type (e.g., drones with drones, observers with observers). These results show that sharing parameters for all agents does not improve performance as different strategies and state processing need to be learned within the same policy. 
However, we believe that more advanced sharing strategies can be designed, and that HeMAC provides a robust testbed for research in this field.

The results presented in this section, leveraging experiments with state-of-the-art MARL algorithms, demonstrate that HeMAC is a suitable testbed for the benchmark of such approaches. Furthermore, the level of performance observed with those algorithms clearly indicates their lack of reliability to tackle Heterogeneous Multi-Agent problems and the need to explore further HeMARL approaches.

\section{Conclusion and Future Work}

We proposed HeMAC as a set of benchmark challenges for heterogeneous multi-agent RL to establish a standardized testbed and facilitate research in this field. HeMAC is based on the Gymnasium standard and uses PettingZoo's AEC API to model interactions of agents with various action and sensing capabilities, and offers multiple challenges allowing users to control both complexity and heterogeneity between agents. The environment focuses on role specialization and cooperative interaction within a team of agents, and we propose 11 scenarios with increasing complexity. HeMAC is designed to advance the development of innovative methods within HeMARL. Additionally, it is compatible with traditional multi-agent reinforcement learning (MARL) and single-agent reinforcement learning (RL), enabling straightforward comparisons with state-of-the-art approaches. We performed an initial benchmark of state-of-the-art MARL algorithms such as MAPPO and QMIX, highlighting that HeMAC is a convenient testbed for the benchmark of MARL algorithms, exploring a range of scenarios in growing complexity and allowing a measure of the performance and limitations of each approach tested. Lastly, the results obtained with respect to the level of performance of the benchmarked state-of-the-art algorithms demonstrate the need for further research in HeMARL. 

In future work, we aim to enhance HeMAC by incorporating a broader range of scenarios and agent types to further increase heterogeneity and complexity, thereby providing a more comprehensive set of benchmarking problems. Additionally, we plan to integrate training and evaluations of both existing and novel heterogeneous multi-agent reinforcement learning (such as HARL) algorithms using HeMAC as a standardized testing platform. Furthermore, we open-source the code \cite{hemac} to enable users to create their own new scenarios and types of agents with new capabilities, observation spaces, and action spaces as required. To facilitate this, we provide a standardized basic agent that serves as a foundation for all other agents, defining the minimum requirements for creating new ones. We welcome contributions of additional scenarios and agents from the community and aim to establish HeMAC as a standard benchmark for evaluating progress on HeMARL problems.







\end{document}